# Spin-Current Shot Noise in Mesoscopic Conductors

Yuhui He

Department of Microelectronics, Peking University, Beijing 100871, China

**Abstract**: In this paper, we present a method to investigate the spin-current shot noise in mesoscopic conductors, by using scattering matrix theory and Green's function technique. We first derive a general expression for the spin-current noise at zero-frequency limit, and extract the shot-noise component by discussing it at zero-temperature limit. The expression indicates that the spin-current shot noise in one lead is caused by the transmissions to the spin-resolved states in this lead and the interferences of these transmissions. As an application, we simulate the spin-current shot noise in a spin transistor, and discuss its dependence on the device parameters and the bias voltages applied to the transistor. The knowledge we gain from this study will help researchers to evaluate the spin-current shot noise in the future spintronic devices.

# I. Introduction

As the dimensions of semiconductor devices go on scaling down, the device performance becomes more sensitive to the noise. In this aspect, the charge-current noise in nanoscale devices has been investigated extensively.[1,2] However, the spin-current noise, which characterizes the correlation of spin-currents, has received much less attention up to now.[3,4] For spintronic devices such as spin transistors [5] it is necessary and beneficial to investigate the spin-current noise. In these devices, the spin currents rather than the charge currents are used as the carrier of information. Therefore, to fully evaluate the performance of spintronic devices, knowledge of the spin-current noise is essential.

Generally speaking, there are two kinds of intrinsic noise in conductors.[6] One is the thermal noise, also called Nyquist-Johnson noise, which is caused by the thermal motion of particles in conductors. The other is the shot noise, caused by the discreteness of particles. In this work we focus on the spin-current shot noise in mesoscopic conductors, since it is quite different from its counterpart in the classical limit.

Theoretically, a few methods have been developed on this problem, including a semiclassical theory based on an extended spin-dependent Boltzman-Langevin equation [3] and a full quantum mechanical treatment using nonequilibrium Green's function technique [4]. Here we propose our approach based on the scattering-matrix (S-matrix) theory and Green's function technique. Compared to other approaches, S-matrix theory is the most successful and widely accepted theory for the electron transport in mesoscopic conductors, while Green's function technique is a powerful tool for calculation and numerical simulation.[7]

In section II, we demonstrate our theoretical approach to the spin-current shot noise in mesoscopic conductors, and illustrate the underlying physics in our final expression for the shot noise.

As an application, we calculate the spin-current shot noise in a spin transistor (spin field effect transistor) in section III. Applying our general result to this particular structure, we obtain the expression for the spin-current shot noise in one lead. Then we employ recursive Green's function technique and surface Green's function technique to facilitate our calculation. After that, numerical results for the spin-current shot noise are presented, and its dependences on the device parameters and bias voltages are further discussed.

# II. A general derivation for spin-current shot noise

We consider a general mesoscopic system: a central conductor with multiple two-dimensional (2D) leads connected to the corresponding electrodes.

We start from the definition of the spin-current noise in $\alpha$ lead:

$$S_\alpha(t,t') = \frac{1}{2} < \Delta P_\alpha(t)\Delta P_\alpha(t') + \Delta P_\alpha(t')\Delta P_\alpha(t) >, \qquad (1.1)$$

where $\Delta P_\alpha(t) = P_\alpha(t) - <P_\alpha(t)>$ is the fluctuation of spin current in that lead. Here the spin-current in $\alpha$ lead is defined as $P_\alpha(t) = I_{\alpha\uparrow}(t) - I_{\alpha\downarrow}(t)$, where $I_{\alpha\sigma}(t)$ is the spin-resolved electrical current with spin $\sigma$. Then we arrive at the following expression for the noise

$$S_\alpha(t,t') = S_\alpha^{\uparrow\uparrow}(t,t') + S_\alpha^{\downarrow\downarrow}(t,t') - S_\alpha^{\uparrow\downarrow}(t,t') - S_\alpha^{\downarrow\uparrow}(t,t'), \qquad (1.2)$$

where $S_\alpha^{\sigma\sigma'}(t, t')$ is the spin-resolved current noise: [8]

$$S_\alpha^{\sigma\sigma'}(t,t') = <\Delta I_{\alpha\sigma}(t)\Delta I_{\alpha\sigma'}(t') + \Delta I_{\alpha\sigma'}(t')\Delta I_{\alpha\sigma}(t)>. \qquad (1.3)$$

In a similar way, we can derive the charge-current noise in α lead $C_\alpha(t, t')$ as

$$C_\alpha(t,t') = S_\alpha^{\uparrow\uparrow}(t,t') + S_\alpha^{\downarrow\downarrow}(t,t') + S_\alpha^{\uparrow\downarrow}(t,t') + S_\alpha^{\downarrow\uparrow}(t,t'). \tag{1.4}$$

Eq. (1.2) and (1.4) show that the difference between spin-current noise and charge-current noise lies in the term $S_\alpha^{\sigma\,-\sigma}$, which denotes *time-domain correlation of fluctuation of spin-opposite currents* $I_{\alpha\sigma}$ and $I_{\alpha-\sigma}$. Thus we will pay particular attention to it in the discussion and simulation.

At steady state, the average currents have no ac components, that is, $<I(\omega)>=0$ for $\omega\neq 0$. With this in mind, we transfer Eq. (1.3) into the frequency domain,

$$2\pi\delta(\omega+\omega')S_\alpha^{\sigma\sigma'}(\omega) = \frac{1}{2}<I_{\alpha\sigma}(\omega)I_{\alpha\sigma'}(\omega') + I_{\alpha\sigma'}(\omega')I_{\alpha\sigma}(\omega)>. \tag{1.5}$$

To discuss the spin-resolved current operator $I_{\alpha\sigma}$, we use second quantization and introduce operators $a^\dagger_{\alpha m\sigma}(\varepsilon)$ and $a_{\alpha m\sigma}(\varepsilon)$, which create and annihilate incident electrons with energy $\varepsilon$ and spin $\sigma$ at the $m$ channel in $\alpha$ lead. Next, we introduce $b^\dagger_{\alpha m\sigma}(\varepsilon)$ and $b_{\alpha m\sigma}(\varepsilon)$, which create and annihilate electrons in the outgoing states at the same channel in that lead. The two sets of operators satisfy the same commutation rules and are related by the S-matrix of the conductor: [9]

$$b_{\beta n\sigma'} = \sum_{\alpha m\sigma} s_{\beta n\sigma',\alpha m\sigma} a_{\alpha m\sigma}. \tag{1.6}$$

With the help of some interim quantities like the scattering states and field operators, we get the expression for the spin-resolved current operator $I_{\alpha\sigma}$ at the low-frequency limit ($\omega \leq 10^{12}$Hz, and a more detailed derivation can be found in [10]):

$$I_{\alpha\sigma}(\omega) = e\sum_m \int d\varepsilon [a^+_{\alpha m\sigma}(\varepsilon)a_{\alpha m\sigma}(\varepsilon+\hbar\omega) - b^+_{\alpha m\sigma}(\varepsilon)b_{\alpha m\sigma}(\varepsilon+\hbar\omega)]. \tag{1.7}$$

To get the ensemble averages of the above operators and the correlations of them, we introduce an important relation based on the assumption of reflection-less contact [11] before proceeding on:

$$<a^+_{\alpha m\sigma}(\varepsilon)a_{\beta n\sigma'}(\varepsilon')> = \delta_{\alpha m\sigma,\beta n\sigma'}\delta(\varepsilon-\varepsilon')f_\alpha(\varepsilon). \tag{1.8}$$

Here $f_\alpha$ is the Fermi distribution function of $\alpha$ lead. The equation tells that the incident electrons in $\alpha$ lead obey equilibrium Fermi distribution of the corresponding $\alpha$ electrode.

Based on this relation, the ensemble average of the spin-resolved current can be written as

$$<I_{\alpha\sigma}(t)> = \frac{e}{h}\sum_m \int d\varepsilon(f_\alpha - \sum_{\beta n\sigma'} T_{\alpha m\sigma,\beta n\sigma'} f_\beta), \tag{1.9}$$

where $T_{\alpha m\sigma,\beta n\sigma'} = |s_{\alpha m\sigma,\beta n\sigma'}|^2$. Since S-matrix is unitary, we have $\sum_{\beta n\sigma'} T_{\alpha m\sigma,\beta n\sigma'} = 1$. Inserting it into the above equation, we obtain the Landauer-Büttiker type expression for the spin-resolved currents:

$$<I_{\alpha\sigma}(t)> = \frac{e}{h}\int d\varepsilon \sum_{\beta\sigma'} T_{\alpha\sigma,\beta\sigma'}(\varepsilon)[f_\alpha(\varepsilon) - f_\beta(\varepsilon)]. \tag{1.10}$$

Here $T_{\alpha\sigma,\beta\sigma'} = Tr(s^\dagger_{\alpha\sigma,\beta\sigma'} s_{\alpha\sigma,\beta\sigma'}) = \sum_{m,n} Tr(s^*_{\alpha n\sigma,\beta m\sigma'} s_{\alpha n\sigma,\beta m\sigma'})$ is the spin-resolved transmission probability from all states with spin $\sigma'$ in $\beta$ lead to all electron states with spin $\sigma$ in $\alpha$ lead.

Using Eq. (1.6) and (1.8), we also get the following expression for $S_\alpha^{\sigma\sigma'}$

$$S_\alpha^{\sigma\sigma'}(\omega) = \frac{e^2}{2h}\sum_{\beta\gamma}\int d\varepsilon \sum_{m,m'} Tr[A_{\beta\gamma}(\alpha m\sigma,\varepsilon,\varepsilon+\hbar\omega)A_{\gamma\beta}(\alpha m'\sigma',\varepsilon+\hbar\omega,\varepsilon)][f_\beta(1-f_r^+) + f_r^+(1-f_\beta)], \tag{1.11}$$

where $f_\beta = f_\beta(\varepsilon), f_\gamma^+ = f_\gamma(\varepsilon+\hbar\omega)$, and the elements of matrix $A_{\beta\gamma}(\alpha m\sigma, \varepsilon, \varepsilon')$ are

$$[A_{\beta\gamma}(\alpha m\sigma,\varepsilon,\varepsilon')]_{l\sigma',n\sigma''} = \delta_{\beta l\sigma',\alpha m\sigma}\delta_{\gamma n\sigma'',\alpha m\sigma} - s^*_{\alpha m\sigma,\beta l\sigma'}(\varepsilon)s_{\alpha m\sigma,\gamma n\sigma''}(\varepsilon').$$

Eq. (1.11) is a general expression for the spin-resolved current shot noise spectrum in $\alpha$ lead. First we discuss it at the equilibrium state. Since all the electrodes have the same Fermi levels $f_\alpha = f_\beta = \ldots = f_\gamma$ at equilibrium, we simplify Eq. (1.11) at the zero-frequency limit as follows

$$S_\alpha^{\sigma\sigma'} = \frac{e^2 kT}{h}\int d\varepsilon(-\frac{\partial f}{\partial \varepsilon})[-T_{\alpha\sigma',\alpha\sigma} - T_{\alpha\sigma,\alpha\sigma'} + 2\delta_{\sigma\sigma'}\sum_{\beta\sigma''}T_{\beta\sigma'',\alpha\sigma}]. \quad (1.12)$$

The above equation illustrates that at equilibrium state the spin-resolved current noise at zero-frequency limit is determined by the corresponding spin-resolved transmissions near the equilibrium Fermi level of the system. We further find that if the spin-flip mechanisms cannot be neglected in the system, transmissions between spin-opposite states will occur. That is, $T_{\beta\sigma, \alpha-\sigma} \neq 0$, and so $S_\alpha^{\sigma-\sigma} \neq 0$.

In the following, by utilizing Green's function (GF) technique we will go beyond the equilibrium situation to the steady-state situation. We first write down the full Green's function of central conductor

$$G^r(x,x') = <x|\frac{1}{\varepsilon I - H_C - \Sigma^r}|x'>, \quad (1.13)$$

where $H_C$ is the Hamiltonian of the central conductor, $I$ is a unit matrix and the self-energy $\Sigma^r = \sum_{\alpha\sigma}\Sigma_{\alpha\sigma}^r$ is the sum of couplings to electronic states with spin $\sigma$ in each lead. Here we define the spin-resolved self-energy $\Sigma_{\alpha\sigma}^r$ as

$$\Sigma_{\alpha\sigma}^r(\varepsilon) = \frac{1}{2\pi}\int\frac{\Gamma_{\alpha\sigma}(\varepsilon')d\varepsilon'}{\varepsilon-\varepsilon'+i\eta}, \quad (1.14)$$

where $\Gamma_{\alpha\sigma}$ is the related line-width function.[12]

Using Fisher-Lee relation [13, 14], we express the S-matrix elements in terms of Green's functions

$$s_{\alpha m\sigma,\beta n\sigma'} = \delta_{\alpha m\sigma,\beta n\sigma'} - 2\pi i\sum_{x,x'}w_{\alpha m\sigma,x}G^r(x,x')w_{x',\beta n\sigma'}, \quad (1.15)$$

where $2\pi\sum_m w_{x',\alpha m\sigma}w_{\alpha m\sigma,x} = [\Gamma_{\alpha\sigma}(\varepsilon)]_{x',x}$.

After a lengthy but straightforward calculation, we obtain the expression for the spin-resolved current noise at zero-frequency limit,

$$S_\alpha^{\sigma\sigma'} = \frac{e^2}{2h}\int d\varepsilon \sum_{i=1}^{5} s_{\alpha,i}^{\sigma\sigma'}, \quad (1.16)$$

where

$$s_{\alpha,1}^{\sigma\sigma'} = -Tr(\Gamma_{\alpha\sigma}G^r\Gamma_{\alpha\sigma'}G^r + \Gamma_{\alpha\sigma}G^a\Gamma_{\alpha\sigma'}G^a)(-2\frac{\partial f_\alpha}{\partial \varepsilon})kT,$$

$$s_{\alpha,2}^{\sigma\sigma'} = 2\delta_{\sigma\sigma'}\sum_{\beta\sigma''}Tr(\Gamma_{\alpha\sigma}G^r\Gamma_{\beta\sigma''}G^a)F_{\alpha\beta},$$

$$s_{\alpha,3}^{\sigma\sigma'} = -\sum_{\beta\sigma'',\gamma\sigma'''} Tr(\Gamma_{\alpha\sigma}G^r\Gamma_{\gamma\sigma'''}G^a\Gamma_{\alpha\sigma'}G^r\Gamma_{\beta\sigma''}G^a)F_{\alpha\beta},$$

$$s_{\alpha,4}^{\sigma\sigma'} = -\sum_{\beta\sigma'',\gamma\sigma'''} Tr(\Gamma_{\alpha\sigma}G^r\Gamma_{\beta\sigma''}G^a\Gamma_{\alpha\sigma'}G^r\Gamma_{\gamma\sigma'''}G^a)F_{\alpha\beta},$$

$$s_{\alpha,5}^{\sigma\sigma'} = +\sum_{\beta\sigma'',\gamma\sigma'''} Tr(\Gamma_{\alpha\sigma}G^r\Gamma_{\gamma\sigma'''}G^a\Gamma_{\alpha\sigma'}G^r\Gamma_{\beta\sigma''}G^a)F_{\beta\gamma}.$$

Here $F_{\beta\gamma} = f_\beta(1-f_\gamma) + f_\gamma(1-f_\beta)$. In the above derivation, an relation $i(G^r - G^a) = G^r\Gamma G^a$ in the Green's function theory has been employed many times.

Eq. (1.16) is a general result for the mesoscopic spin-resolved current noise at zero-frequency limit. We have grouped the terms in the equation into five categories, with each group having a different physical significance.

At zero-temperature limit, we can extract the shot-noise component from Eq. (1.16), since at this limit thermal noise will vanish leaving only shot noise. When $kT=0$, $s_{\alpha,1}^{\sigma\sigma'}$ will vanish and so it represents thermal noise. For $s_{\alpha,2}^{\sigma\sigma'}$, $s_{\alpha,3}^{\sigma\sigma'}$, $s_{\alpha,4}^{\sigma\sigma'}$ and $s_{\alpha,5}^{\sigma\sigma'}$, those terms with $\beta=\alpha$, or $\beta=\gamma$ in the sum will also vanish and they represent thermal noise too. The remaining terms in $s_{\alpha,2}^{\sigma\sigma'}$, $s_{\alpha,3}^{\sigma\sigma'}$, $s_{\alpha,4}^{\sigma\sigma'}$ and $s_{\alpha,5}^{\sigma\sigma'}$ represent shot noise, and they are commonly in the form of a product of two factors. First we discuss the physical meaning of the factor $Tr(\Gamma_{\alpha\sigma}G^r\Gamma_{\beta\sigma''}G^a)$ in $s_{\alpha,2}^{\sigma\sigma'}$. We know that the spin-resolved transmission probability $T_{\alpha\sigma,\beta\sigma'}$ can be expressed through the Green's function as $T_{\alpha\sigma,\beta\sigma'} = Tr(s^\dagger_{\alpha\sigma,\beta\sigma'} s_{\alpha\sigma,\beta\sigma'}) = Tr(\Gamma_{\alpha\sigma}G^r\Gamma_{\beta\sigma'}G^a)$. [7]. Thus the factor $Tr(\Gamma_{\alpha\sigma}G^r\Gamma_{\beta\sigma''}G^a)$ in $s_{\alpha,2}^{\sigma\sigma'}$ represents spin-resolved transmission. With the above expression in mind, we also find that the factor such as $Tr(\Gamma_{\alpha\sigma}G^r\Gamma_{\gamma\sigma'''}G^a\Gamma_{\alpha\sigma'}G^r\Gamma_{\beta\sigma''}G^a)$ in $s_{\alpha,3}^{\sigma\sigma'}$, $s_{\alpha,4}^{\sigma\sigma'}$ and $s_{\alpha,5}^{\sigma\sigma'}$ denotes the interference between different transmission processes $T_{\alpha\sigma,\gamma\sigma''}$ and $T_{\alpha\sigma',\beta\sigma''}$. The other factor in the production such as $[f_\beta(1-f_\gamma)+f_\beta(1-f_\gamma)]$ ($\beta\neq\gamma$) actually restrains an energy range between the Fermi levels of electron reservoir $\beta$ and $\gamma$. Only in this energy rangy, transmissions or the interferences of transmissions will contribute to the shot noise. In Fig. 1, we show the interferences of spin-resolved transmissions denoted by $s_{\alpha,3}^{\sigma\sigma'}$, $s_{\alpha,4}^{\sigma\sigma'}$ and $s_{\alpha,5}^{\sigma\sigma'}$ in the corresponding energy ranges.

Since the spin-opposite current noise $S_\alpha^{\sigma,-\sigma}$ makes the difference between spin-current noise and charge-current noise, we further investigate it in details. We insert a complete set of mutually orthogonal states into Eq. (1.16) and find it will involve factors like $G_{\sigma,-\sigma}$, which is the spin-resolved submatrix of Eq. (1.13) and describes spin-flip propagations. If spin-flip process is neglectable, these factors are zero. Otherwise, the spin-opposite current noise will exist and distinguish the spin-current noise from charge-current noise. This conclusion is in accordance with that indicated by Sauret and Feinberg. [8]

To investigate the quantitative dependence of the spin-current shot noise on device parameters and applied bias voltages, we will apply our method to the calculation of spin-current shot noise in a spin transistor, which is a typical kind of spintronic device. Numerical results and discussions are presented in section III.

### III. Spin-current shot noise in a spin transistor

It is discovered that in two dimensional electron gas (2-DEG) made by narrow band gap semiconductor materials, the electron spin will precess under the constraining voltage perpendicular to the 2-DEG plain. This is the so called Rashba effect.[15, 16] Based on this effect, a structure for spin transistor (Datta-Das Transistor) is proposed: when there is spin-polarized current injection from the source electrode made by ferromagnetic materials, we can tune the spin precession in the channel by changing the gate voltages, and detect it at the drain electrode also made by ferromagnetic materials.[5]

In this section, we will investigate the spin-current shot noise in such a spin transistor: 2-DEG lies in the *x-y* plain, and the channel is along the *x* direction. In the z direction, we apply gate voltage to induce Rashba effect. At the left and right ends, there are 2D ferromagnetic leads connected to the corresponding electrodes (as seen in Fig. 2). Here we have set the magnetic polarization of the leads to be parallel.

For the Ferromagnetic leads, the Hamiltonian is written as $H_\alpha = \Sigma_{k\sigma} (\varepsilon_k + \sigma\mu) a^\dagger_{k\sigma} a_{k\sigma}$. And the Hamiltonian of the central 2-DEG is

$$H_C = \frac{p_x^2}{2m} + \frac{p_y^2}{2m} + \frac{\alpha}{\hbar}(\sigma_x p_y - \sigma_y p_x) + U(x,y), \qquad (2.1)$$

where $\alpha \sim <\Psi(z)|(d/dz)V(z)|\Psi(z)>$ is the spin-momentum interaction coefficient. We employ the tight-binding lattice representation $\{|\Phi^m_{n,\sigma}>\}$, where *n* denotes the discrete lattice along the transverse *y* direction, and *m* denotes that along the longitudinal *x* direction. Then we write down $H_C$ layer by layer along *x* direction: $H_m$ (m=*1, 2, ..., M*) and $V_{m,m+1}$ (m=*0, 1, ..., M*). Here $H_m$ is the intra-layer Hamiltonian for the *m* layer, and $V_{m,m+1}$ is the inter-layer Hamiltonian for the couplings between the *m* and *m+1* layer. Their dimension is determined by the number of discrete lattices along *y* direction.

After discretization of Eq. (2.1), the elements of $H_m$ and $V_{m,m+1}$ are as follows

$$\begin{cases} (H_m)_{n\sigma,n\sigma} = 2t_x + 2t_y + U_{m,n} \\ (H_m)_{n\pm1\sigma,n\sigma} = -t_y \\ (H_m)_{n+1\sigma,n\bar{\sigma}} = -i\gamma_y \\ (H_m)_{n-1\sigma,n\bar{\sigma}} = i\gamma_y \end{cases}, \qquad (2.2)$$

$$\begin{cases} (V_{m-1,m})_{n\sigma,n\sigma} = -t_x \\ (V_{m-1,m})_{n\uparrow,n\downarrow} = -\gamma_x \\ (V_{m-1,m})_{n\downarrow,n\uparrow} = \gamma_x \end{cases}, \qquad (2.3)$$

where $t_{x,y} = \hbar^2/2ma^2_{x,y}$, and $\gamma_{x,y} = \alpha/2a_{x,y}$. Here $a_x$ and $a_y$ are the tight-binding lattice parameters along *x* and *y* directions.

Applying Eq. (1.16) to the above structure, we get the spin-resolved current shot noise in the left lead as follows:

$$S_L^{\sigma\sigma'} = \frac{e^2}{h} \int_{\mu\min}^{\mu\max} d\varepsilon Tr(\delta_{\sigma\sigma'}\Gamma_{L\sigma} G^r_{1M} \Gamma_R G^a_{M1} - \Gamma_{L\sigma'} G^r_{1M} \Gamma_R G^a_{M1} \Gamma_{L\sigma} G^r_{1M} \Gamma_R G^a_{M1}), \qquad (2.4)$$

where $\mu_{\max} = \max\{\mu_L, \mu_R\}$ and $\mu_{\min} = \min\{\mu_L, \mu_R\}$. From Eq. (2.4), we need to calculate the full GF $G^r_{1M}$ and $G^a_{M1}$, which denote the electron propagation between the leftmost *1* layer and the rightmost *M* layer. In this work we build up them recursively by employing the lattice Green's function technique.[17]

First, we consider the rightmost *M* layer, and write its *partial* Green's function as

$$g_M^r = [\varepsilon I - H_M - \Sigma_R^r]^{-1}. \tag{2.5}$$

We can see that this Green's function is neither full Green's function nor free Green's function of $M$ layer. It only takes the coupling to its right part into account. Thus we call it *partial* Green's function.

Then, we move on to $M-1$ layer, and calculate its partial GF, again taking into account the coupling to its right parts,

$$g_{M-1}^r = [\varepsilon I - H_{M-1} - V_{M-1,M} g_M^r V_{M,M-1}]^{-1}. \tag{2.6}$$

We move on by adding layer by layer according to the Dyson's equation. For $m$ layer, its partial GF is

$$g_{m-1}^r = [\varepsilon I - H_{m-1} - V_{m-1,m} g_m^r V_{m,m-1}]^{-1}, \tag{2.7}$$

and finally arrive at the leftmost $1$ layer. Now we take the coupling to both left and right side of it into account, and have

$$g_1^r = [\varepsilon I - H_1 - V_{1,2} g_2^r V_{2,1} - \Sigma_L^r]^{-1}. \tag{2.8}$$

Since couplings to all parts of the device have been included, $g_1^r$ is just the full GF of layer 1

$$G_{11}^r = g_1^r. \tag{2.9}$$

To obtain $G_{M1}^r$, we now divide the system into two parts at the interface of $M-1$ layer and $M$ layer. Then, we expand $G_{M1}^r$ according to Dyson equation

$$G_{M1}^r = g_M^r V_{M,M-1} G_{M-1,1}^r, \tag{2.10}$$

where $g_M^r$ is partial Green's function, and $G_{M-1,1}^r$ is the full Green's function.

Similarly, to build up $G_{M-1,1}^r$, we divide the device into two parts at the interface of $M-2$ layer and $M-1$ layer, and expand $G_{M-1,1}^r$ as:

$$G_{M-1,1}^r = g_{M-1}^r V_{M-1,M-2} G_{M-2,1}^r \tag{2.11}$$

Thus we obtain a recursive formula

$$G_{m-1,1}^r = g_{m-1}^r V_{m-1,m-2} G_{m-2,1}^r. \tag{2.12}$$

Combining the above results, we get the expression for $G_{M1}^r$

$$G_{M1}^r = g_M^r V_{M,M-1} g_{M-1}^r V_{M-1,M-2} \cdots V_{2,1} G_{11}^r. \tag{2.13}$$

$G_{1M}^r$ can be calculated in similar way. Compared with other techniques, this recursive approach will obviously relax the memory requirements and enhance the computational speed.

It is worth mentioning that both the above calculations for full GF and for line-width functions in Eq. (2.4) need the surface Green's functions of the leads

$$\Gamma_{L\sigma} = -2\,\text{Im}(V_{1,-1\sigma} g_{-1,\sigma}^r V_{-1\sigma,1}). \tag{2.14}$$

We build up surface GF $g_{-1,\sigma}^r$ with the surface Green's function technique developed by Lee and Joannopoulos [18], because this approach is easier to implement and faster than any other approach currently available.

In the following, we present our simulation results and discuss the underlying physics. First, we show the dependence of the spin current and its shot noise on the channel length in Fig. 3. The tight-binding lattice parameter *a* is set to be *2.8nm*, the spin-momentum interaction coefficient α is $23.8*10^{-12}$ *eV·m*, while the nearest hopping energies $t_x$ and $t_y$ are *0.125eV*.[19] Here we set the source-drain bias $\mu_L - \mu_R = 0.8\ eV$. We alter the number of discrete lattices along the channel length direction from *40* to *240*, and set the number of discrete lattices along the channel width direction to be *10*. This results in a channel length varying from *112nm* to *672 nm*, and a fixed channel width of *28nm*. From the figure, we can see as the length of the channel increases, both the spin-resolved currents $I_{\alpha\sigma}$ and the shot noise $S^{\alpha}_{\sigma\sigma'}$ vary periodically. This can be understood in the following way: given a fixed gate voltage, the angular frequency of spin precession in the channel of spin transistor is determined. As the channel length gradually increases, the amount of spin rolling from source to drain will change periodically (as seen in Fig. 2), thus the spin-current and its shot noise will also change periodically. Besides, we find that the magnitude of the spin-opposite current shot noise $S_L^{\sigma\ -\sigma}$ is much smaller than the spin-parallel current shot noise $S_L^{\sigma\sigma}$. We think there are two factors contributing to this difference. First, $S_L^{\sigma\sigma}$ contains a spin-resolved transmission term while $S_L^{\sigma\ -\sigma}$ does not. Second, from Eq. (2.4) we find that $S_L^{\sigma\ -\sigma}$ is a product of transmissions to spin-opposite (σ and −σ) states in *L* lead. These transmissions are repulsive, because the sum of them should be no greater than 1. Therefore, the product of them is much smaller than themselves. When the difference between these transmissions is the largest, the product of them is smallest while the spin current is largest. This explains why $S_L^{\sigma\ -\sigma}$ is minimal at the extreme of the spin current. Another interesting phenomenon is that the spin-opposite current shot noise is always negative. The reason can be found in Eq. (2.4): There is a "−" sign before the product of the transmissions to spin-opposite states in the expression for $S_L^{\sigma\ -\sigma}$.

Then, the spin-current shot noise $S_L$, its spin-opposite component $S_L^{\sigma\ -\sigma}$ and its Fano factor as a function of the channel width are shown in Fig. 4. Fano factor, which characterize the mesoscopic deviation of shot noise from its classical value *2eI*, is defined as [20]

$$\gamma \equiv \frac{<(\Delta I)^2>}{2eI}. \tag{2.17}$$

In this figure, we set *M=94*, at which length the spin current has extreme value according to Fig. 3. We find that both the spin current (unshown) and its shot noise increase linearly with the channel width. We attribute this phenomenon to the linearly increased number of one-dimensional (1D) subbands with wider transistor channel in the bias-restrained energy range. From the lower figure, we can see the Fano factor is always lesser than 1. This means that the spin-current shot noise in mesoscopic conductors is lesser than their classical counterparts. This suppression of shot noise is caused by Pauli exclusion principle [12].

Fig. 5 shows the dependence of spin-current shot noise $S_L$, one of its components $S_L^{\sigma\ -\sigma}$ and its Fano factor on the spin-momentum interaction coefficient α. From (2.1), we can see gradually changed α leads to gradually changed spin-precession angular frequency ω in the channel. Given fixed channel length, this will result in gradually changed amount of spin rolling from the source to drain. Therefore, the spin current (unshown) and its shot noise $S_L$ will oscillate with $\alpha$, but the average values of them change little. From the upper figure, we also find that the spin-opposite current noise $S_L^{\sigma\,-\sigma}$ is nearly zero when $\alpha$ is very small, which meets with our predictions in section II. The lower figure shows that the Fano factor $\gamma_L$ changes roughly in accordance with the $S_L$. This implies that the denominator of $\gamma_L$, that is, the electrical current in the left lead $I_L=I_{L\uparrow}+I_{L\downarrow}$ changes little with the spin-momentum interaction coefficient. Physically, it is because the variance of electrical current $I_L$ as the spin-momentum interaction coefficient changes is quite small compared to the average value of $I_L$.

Fig. 6 shows the spin-current shot noise $S_L$, one of its components $S_L^{\sigma\,-\sigma}$ and the corresponding Fano factor $\gamma_L$ as a function of the source-drain bias. From the upper figure we find that the shot noise is near zero when the source-drain bias is small. We infer that under this condition, few electron states exist within this narrow "energy window", so the current and noise approach to zero. At small bias (less than 0.1eV), $S_L$ and $S_L^{\sigma\,-\sigma}$ increase linearly with the bias. When the bias goes on increasing, however, their changing trends deviate from the linear increment. We attribute this phenomenon to the nonlinearity of the leads' energy band with larger bias. Finally, when the bias is up to *1eV*, the spin-current noise stops increasing and becomes saturated. This is caused by the decreased spin-polarization as the source-drain bias increases. The spin polarization in $\alpha$ lead is defined as $p_\alpha=(I_{\alpha\uparrow}-I_{\alpha\downarrow})/(I_{\alpha\uparrow}+I_{\alpha\downarrow})$. From the inset figure which plots the spin polarization in the left lead $p_L$ as a function of the bias, we find that as the bias increases, the spin-polarization decreases to zero rapidly. This will lead to zero spin current and saturated spin-current shot noise. In the lower figure, we can see the Fano factor of spin-current shot noise decreases monotonously as the bias increases. This means that the increasing speed of spin-current noise in mesoscopic conductor is always lesser than that at the classical limit as the source-drain bias increases. We think this universal suppression of the mesoscopic shot noise is a typical quantum effect due to Pauli exclusion principle.

Finally, for the experimental observation and verification, we suggest the spin noise spectroscopy detection technique developed by Oestreich etc.[22]

## IV. CONCLUSION

In this work, we have developed a general expression for the mesoscopic spin-current noise at zero-frequency limit. The framework is based on scattering matrix theory and Green's function technique. The shot-noise component is extracted at zero-temperature limit, and recognized as a sum of spin-resolved transmissions and interferences of these transmissions, in the corresponding bias-restrained energy ranges. As an example, we apply our method to the calculation of spin-current shot noise in a spin transistor and present our numerical results. We demonstrate that not only is Green's function technique useful in the derivation of a general expression for the spin-current shot noise, but it will also greatly facilitate the calculation and numerical simulation of a particular structure. Our numerical results show that the spin-current shot noise will change roughly in accordance with the spin-current as the channel length and width of the spin transistor changes. It will oscillate as the spin-momentum interaction coefficient varies in the channel. Besides, it will become saturated at large source-drain bias. The knowledge we gain in this work will help us evaluate the spin-current shot noise in mesoscopic conductors and optimize the design of future spintronic devices.

## Acknowledgement

The authors would like to thank Dr. Chun Fan and Aidong Sun for their help in our numerical calculations. All simulations were performed on IBM RS/6000 SP3 at the Center for Computational Science and Engineering at Peking University.

**Figure Captions**

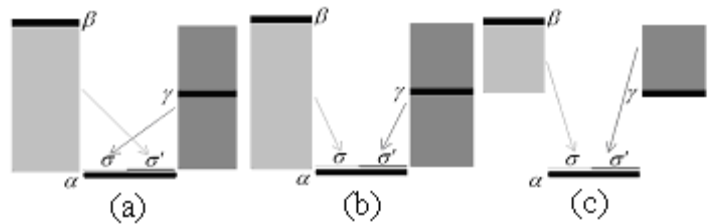

Fig.1. (a), (b), (c) give a schematic view of interference between different transmission processes which lead to spin-current shot noise in $s_{\alpha 3}^{\sigma \sigma'}$, $s_{\alpha 4}^{\sigma \sigma'}$ and $s_{\alpha 5}^{\sigma \sigma'}$, respectively.

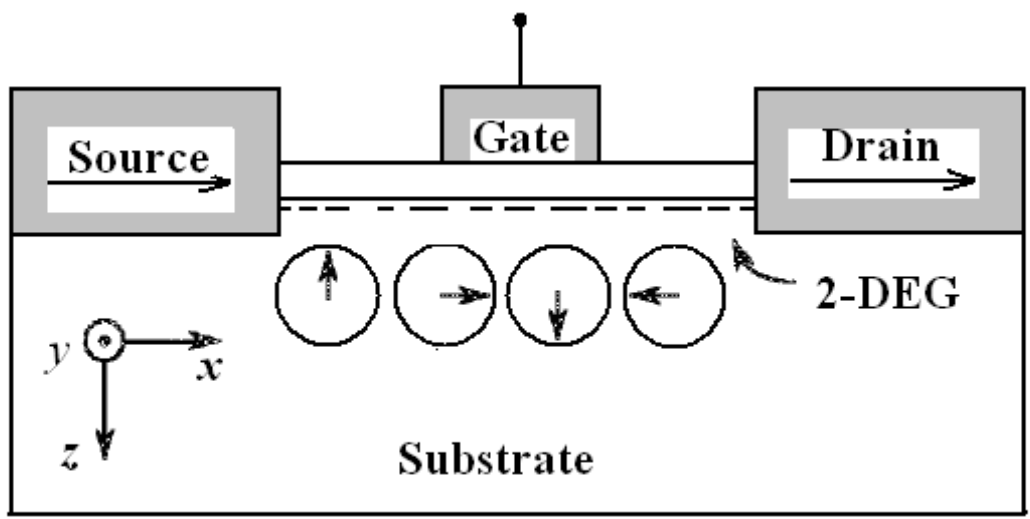

Fig. 2. A sketch map of the spin transistor.

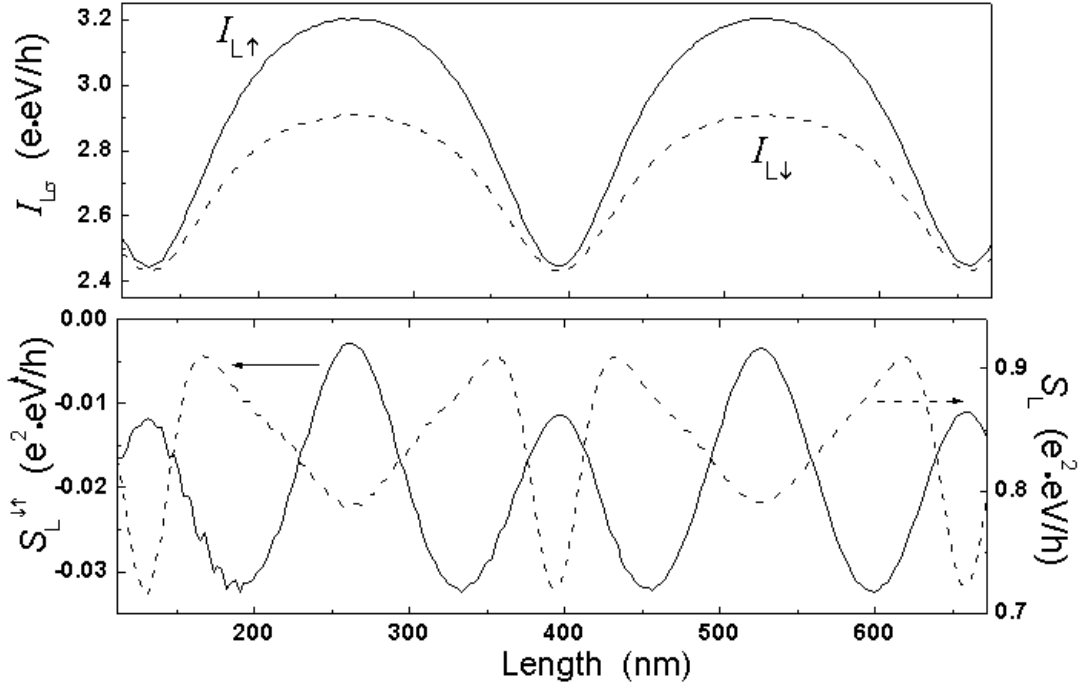

Fig. 3. Spin-resolved currents and the spin-current shot noise as a function of the channel length. Here we have set $N=10$ so that the channel width is about $28.4nm$. $M$ is changed from 40 to 240. $S_L^{\uparrow\downarrow}$ is the spin-opposite current shot noise in the left lead, while $S_L$ is the spin-current shot noise in that lead.

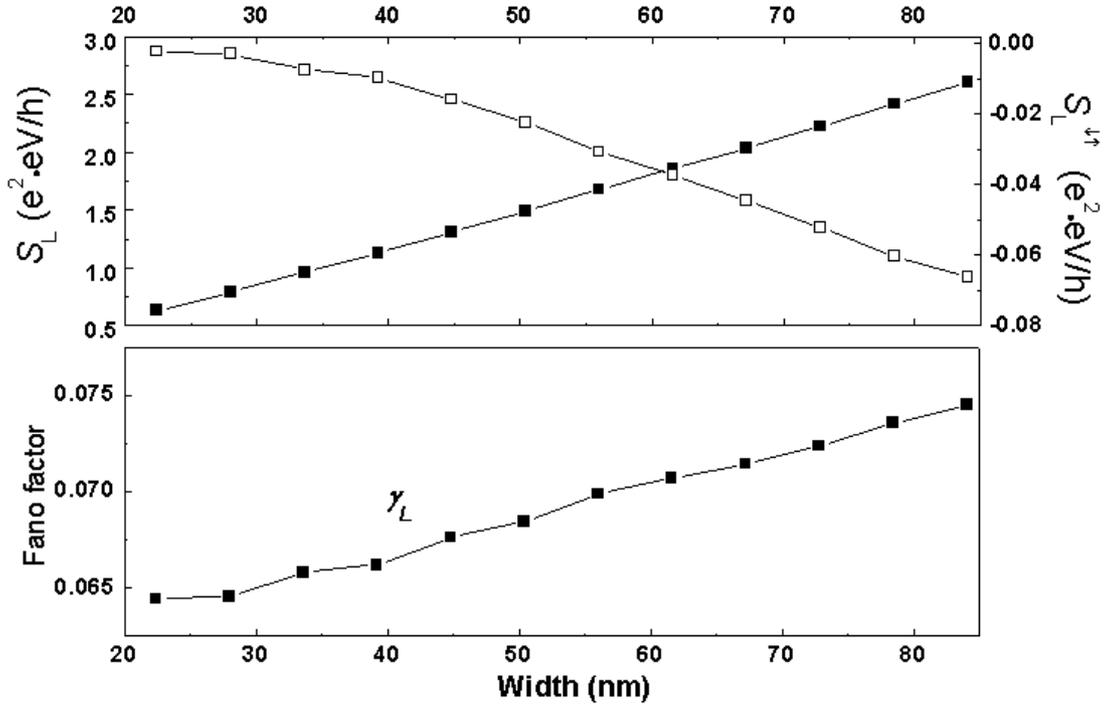

Fig. 4. The spin-current shot noise and its Fano factor as a function of the channel width. $M=94$, while other parameters are the same as Fig. 3. Here $\gamma_L$ is the Fano factor for the spin-current shot noise.

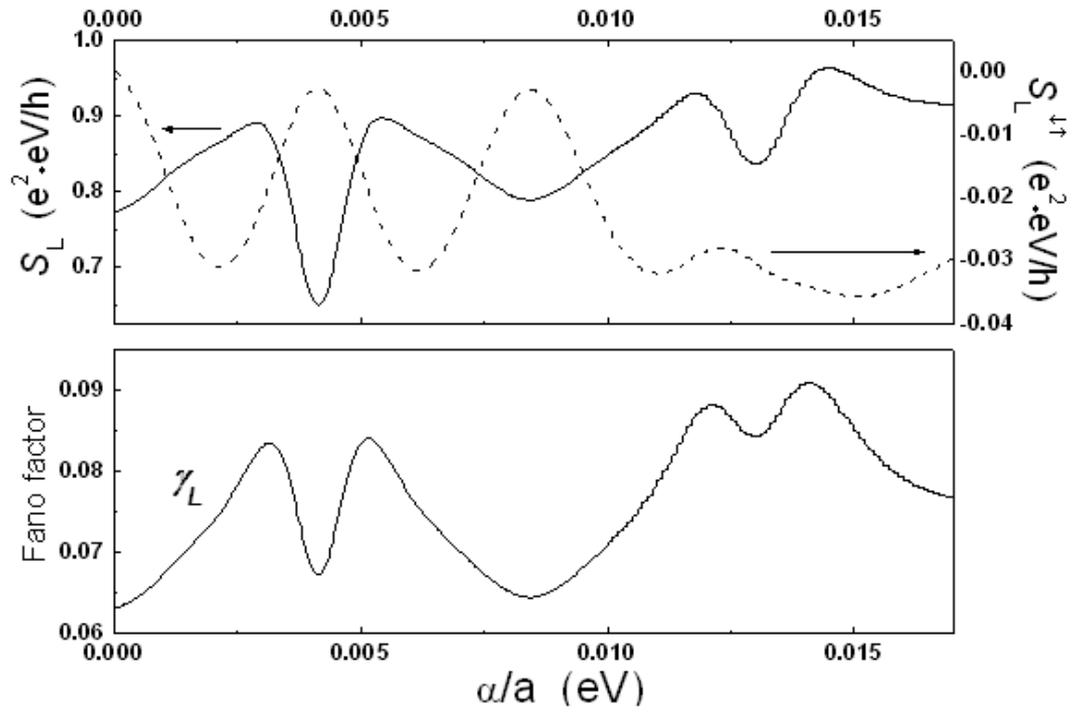

Fig. 5 The spin-current shot noise and its Fano factor as a function of the spin-momentum interaction coefficient $\alpha$. $M=94$, while other parameters are the same as Fig. 3. Here $\gamma_L$ is the Fano factor for the spin-current shot noise.

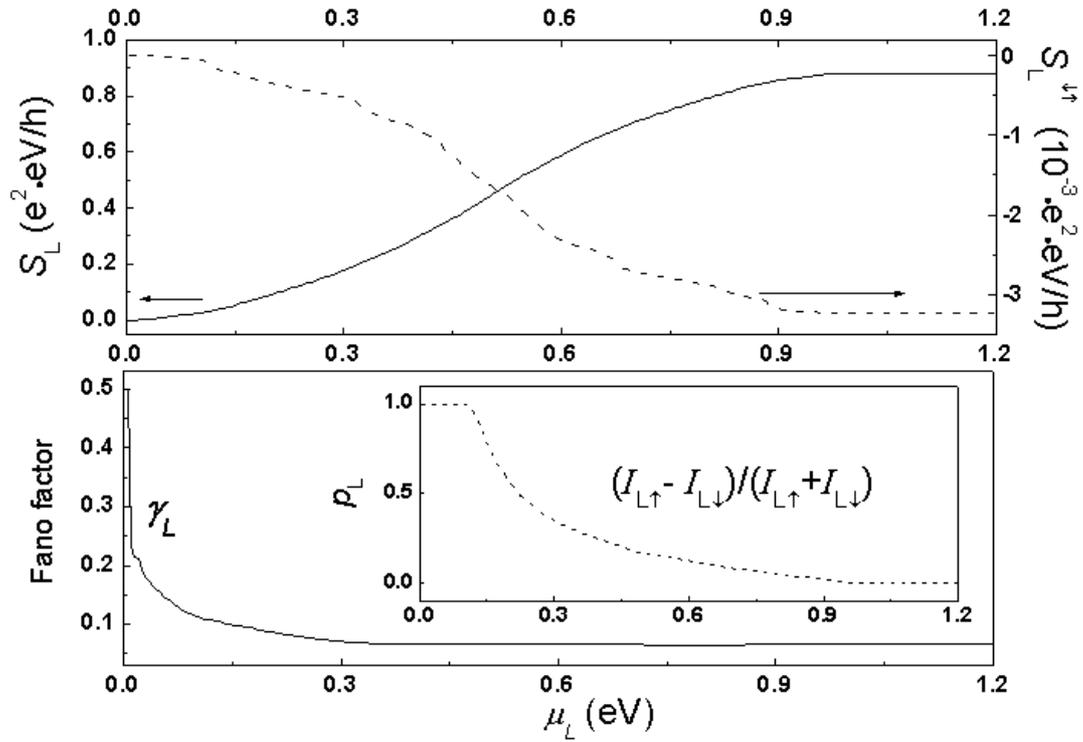

Fig. 6 The spin-current shot noise and its Fano factor as a function of the source-drain bias. Inset is the spin-polarization as a function of the source-drain bias. $M=94$, $N=10$, while other parameters are the same as Fig. 3. $\mu_R$ is set to be zero, and we change the $\mu_L$.